\documentclass[11pt]{article}
\usepackage{amssymb,amsmath,mathrsfs,enumerate}
\usepackage{graphicx,rotate,multicol}
\usepackage[margin=10pt,labelfont=bf]{caption}
\usepackage{cite}
\usepackage[utf8]{inputenc}

\usepackage{wrapfig}

\usepackage[colorlinks=true,
            linkcolor=red,
            urlcolor=blue,
        citecolor=blue]{hyperref}

\usepackage{color}
\usepackage{lineno}
\usepackage{tcolorbox}
\long\def\rpl#1!!!#2!!!{\color[rgb]{.7,0,0}{#1} \color{blue}{#2} \color{black}}

\newsavebox\curwrapfig
\makeatletter
\long\def\wrapfiguresafe#1#2#3{%
  \sbox\curwrapfig{#3}%
  \par\penalty-100%
  \begingroup 
    \dimen@\pagegoal \advance\dimen@-\pagetotal 
    \advance\dimen@-\baselineskip 
    \ifdim \ht\curwrapfig>\dimen@ 
      \break%
    \fi%
  \endgroup%
  \begin{wrapfigure}{#1}{#2}%
    \usebox\curwrapfig%
  \end{wrapfigure}%
}
\makeatother

\let\tilde=\widetilde

\let\bar=\overline

\newcommand{\secant}{{\rm sec}} 
\newcommand{\cosec}{{\rm csc}}

\def \order(#1){{\mathcal O} \left(#1 \right)}

\evensidemargin=\oddsidemargin
\def\Eqn#1{Eq.\ (\ref{#1})}
\def\Eqs#1#2{Eqs.\ (\ref{#1}) and (\ref{#2})}
\allowdisplaybreaks

\title	{\Large\bf
Diluting quark flavor hierarchies using dihedral symmetry
}

\author {\sf Ayushi Srivastava,$^{a,}$\footnote{srivastavaayushi860@gmail.com}
	\quad Miguel Levy,$^{b,}$\footnote{miguelplevy@ist.utl.pt}
	  \quad  Dipankar	Das\,$^{a,}$\footnote{d.das@iiti.ac.in}
   \\[10pt]
\small\em $^a$Department of Physics, Indian Institute of Technology (Indore), Khandwa Road, Simrol, 453552 Indore, India\\ 
\small\em  $^b$Centro de F\'{i}sica Te\'{o}rica de Part\'{i}culas-CFTP and Departamento de F\'{i}sica, Instituto Superior T\'{e}cnico,\\
\small\em      Universidade de Lisboa, Av Rovisco Pais, 1, P-1049-001 Lisboa, Portugal\\ }

\date{}

\begin{document}


\maketitle


\begin{abstract}
We present a $D_4$ flavored extension of the SM which provides an intuitive reasoning for the masses and mixing patterns in the
quark sector. In our model, the Cabibbo mixing angle  
becomes related to the ratio of two vacuum expectation values. In fact, the orders of magnitude of the CKM matrix elements are readily obtained from  
the hierarchical nature of the vacuum expectation values. Moreover, we also show that the smallness of the off-Cabibbo elements
in the CKM matrix is strongly connected to the heaviness of the third generation of quarks.
\end{abstract}

\bigskip
%
The Standard Model~(SM) successfully explains the mechanism responsible for the fermion masses but does not justify
them. The arbitrariness of the Yukawa couplings makes the SM adaptable to any spectrum of fermion masses and mixings
brought in by the experimental measurements. As it happens, the observed quark masses span five orders of magnitudes,
with the third generation of quarks being much heavier than the first two. Furthermore, the quark mixings obey the
following hierarchical pattern\cite{Wolfenstein:1983yz}:
\begin{eqnarray}
\label{e:wolf}
	V \approx \begin{pmatrix}
	1-\lambda^2/2 & -\lambda & \order(\lambda^3) \\
	\lambda & 1-\lambda^2/2 & \order(\lambda^2) \\
	\order(\lambda^3) & \order(\lambda^2) & 1
	\end{pmatrix}\,,
\end{eqnarray}
where $\lambda\approx 0.22$ is the Cabibbo mixing parameter and the matrix, $V$, is known as the Cabibbo-Kobayashi-Maskawa
(CKM) matrix\cite{Cabibbo:1963yz,Glashow:1970gm,Kobayashi:1973fv}. Within the ambit of SM, such hierarchies can only 
originate from some conspiracies within the Yukawa
couplings themselves\cite{Botella:2016krk}. This is part of the problem that is usually dubbed in the literature as the `flavor puzzle'\cite{Feruglio:2015jfa, 
Xing:2014sja, Xing:2020ijf}.
This aspect of the SM, for decades, has fueled speculations that there might exist a deeper theoretical framework which
can offer a more natural insight into the flavor structure.
This article presents an extension beyond the SM~(BSM) with a $D_4$ symmetry, which can make the
quark flavor structure appear more instinctive (for other works on flavor models using $D_4$ symmetry, see \cite{Adulpravitchai:2008yp,
Ishimori:2008gp,Hagedorn:2010mq,Meloni:2011cc, Vien:2014soa, Vien:2013zra, Vien:2018imw, CarcamoHernandez:2020ney, Bonilla:2020hct}). 
An essential ingredient of our model is that the primary sources of masses
for the third generation of quarks have been disentangled from those for the first two generations of
quarks. The hierarchies in the quark masses and mixings are then chiefly attributed to the hierarchies in the vacuum
expectation values~(VEVs) of the different scalar fields. This allows us to relax the Yukawa hierarchies 
in the quark sector considerably along with some new and interesting implications for the CKM matrix. 
We will describe our model in detail in the upcoming sections.

%
We start by laying out some of the basics of $D_4$ symmetry\cite{Ishimori:2010au}.
The discrete group $D_4$ has five irreducible representations which we label
as $\mathbf{1}_{++}$, $\mathbf{1}_{--}$, $\mathbf{1}_{-+}$, $\mathbf{1}_{+-}$, and $\mathbf{2}$. 
For the two-dimensional representation of $D_4$, we opt to work in a basis 
 in which the generators of $D_4$ are given by
\begin{eqnarray}
\label{e:basis}
a= \begin{bmatrix}
0 & -1\\ 1 & 0
\end{bmatrix}, \hspace{1cm} b= \begin{bmatrix}
1 & 0 \\ 0 & -1
\end{bmatrix} \,,
\end{eqnarray}
where $a$ is of order 4 and $b$ is of order 2. 
In this basis, the relevant tensor products in the explicit component form are given by\cite{Das:2019itj}
\begin{subequations}
	\label{e:D4tensors}
	\begin{eqnarray}
	{\begin{bmatrix}x_1 \\ x_2 \end{bmatrix}}_\mathbf{2} \otimes {\begin{bmatrix} y_1 \\ y_2 \end{bmatrix}}_\mathbf{2} &=& 
	{\begin{bmatrix} x_1 y_1 + x_2 y_2 \end{bmatrix}}_{\mathbf{1}_{++}}
	\oplus
	{\begin{bmatrix} x_1 y_2 - x_2 y_1 \end{bmatrix}}_{\mathbf{1}_{--}} \nonumber \\
	&& \oplus
	{\begin{bmatrix}x_1 y_2 + x_2 y_1 \end{bmatrix}}_{\mathbf{1}_{-+}}
	\oplus
	{\begin{bmatrix}x_1 y_1 - x_2 y_2 \end{bmatrix}}_{\mathbf{1}_{+-}} \,,
	\\
	\mathbf{1}_{r,s} \otimes \mathbf{1}_{r',s'}&=&\mathbf{1}_{r\cdot r',s\cdot s'} \,.
	\end{eqnarray}
\end{subequations}

%
Now we will specify the $D_4$ transformations of the different fields in our model. The $i$-th generation of left-handed quark
 doublet is denoted by $Q_{iL}\equiv (p_{iL},n_{iL})^T$. The right-handed charged quark singlets are denoted by $p_{iR}$ and $n_{iR}$ in the up
 and down sectors, respectively. We have four scalar doublets in our model, which we symbolize as $\phi_1$, $\phi_2$,
 $\phi_u$ and $\phi_d$. These fields are assumed to transform under the $D_4$ symmetry as follows:
 \begin{subequations}
 	\label{e:trans}
 	\begin{eqnarray}
 	\label{e:trans1}
 	&& \textbf{2}: \begin{bmatrix} Q_{1L} \\ Q_{2L}\end{bmatrix}
 	\,,\begin{bmatrix} \phi_{1} \\ \phi_{2} \end{bmatrix}
 	\,, \\
 	\label{e:trans2}
 	&& \textbf{1}_{++}: n_{1R} \,,\quad  \textbf{1}_{--}: n_{2R}\,, n_{3R}\,, \phi_u \,,\quad  \textbf{1}_{-+}: p_{2R}\,, p_{3R} \,, \phi_d \,,\quad \textbf{1}_{+-}: Q_{3L} \,, p_{1R} \,.
 	\end{eqnarray} 
 \end{subequations}
As we will see shortly, because of the above transformations, $\phi_u$ and $\phi_d$ will couple exclusively to the up and
down type quarks respectively, which justifies their labeling.
The gauge and $D_4$ invariant Yukawa Lagrangians in the up and down quark sectors are then given by
\begin{subequations}
\label{e:Lag}
	\begin{eqnarray}
	\label{e:Lag2}
	-{\mathscr{L}}_u &=& A_u(\bar{Q}_{1L} \tilde{\phi}_1 - \bar{Q}_{2L} \tilde{\phi}_2 )p_{1R} + B_u(\bar{Q}_{1L} \tilde{\phi}_2 + \bar{Q}_{2L} \tilde{\phi}_1 ) p_{2R} + C_u(\bar{Q}_{1L} \tilde{\phi}_2 + \bar{Q}_{2L} \tilde{\phi}_1 ) p_{3R} \nonumber \\
	&& + X_u \bar{Q}_{3L} \phi_u p_{2R}+ Y_u \bar{Q}_{3L} \phi_u p_{3R} \,, \\
	\label{e:Lag1}
	-{\mathscr{L}}_d &=& A_d(\bar{Q}_{1L} \phi_1 + \bar{Q}_{2L} \phi_2 )n_{1R} + B_d(\bar{Q}_{1L} \phi_2 - \bar{Q}_{2L}\phi_1)n_{2R} 
	+ C_d(\bar{Q}_{1L} \phi_2 - \bar{Q}_{2L}\phi_1)n_{3R} \nonumber \\ 
	&& + X_d \bar{Q}_{3L} \phi_d n_{2R} + Y_d \bar{Q}_{3L} \phi_d n_{3R} \,,
	\end{eqnarray}
\end{subequations}
where $\tilde{\phi}_k = i\sigma_2\phi_k^\star$ with $\sigma_2$ being the second Pauli matrix. For an intuitive understanding
of the upcoming results, we will assume the Yukawa parameters to be real. As such, we will not deliberate so much on the complex phase of the
CKM matrix. We will treat the phase as an independent parameter which, as we have checked, can be easily accommodated by allowing the Yukawa couplings
to be complex.
The mass matrices in the up and down sector that transpire from \Eqn{e:Lag} are 
\begin{eqnarray}
\label{e:massmat}
M_u = \begin{pmatrix}
       	A_u v_1 & B_u v_2 & C_u v_2 \\
       	-A_u v_2 & B_u v_1 & C_u v_1 \\
       	0 & X_u v_u & Y_u v_u
\end{pmatrix} \,,
\quad M_d = \begin{pmatrix}
	  	A_d v_1 & B_d v_2 & C_d v_2  \\
	  	A_d v_2 & -B_d v_1 & -C_d v_1 \\
	  	0 & X_d v_d & Y_d v_d
\end{pmatrix},
\end{eqnarray}
where $v_1$, $v_2$, $v_u$ and $v_d$ represents the VEVs of $\phi_1$, $\phi_2$, $\phi_u$ and $\phi_d$ respectively with the
total electroweak VEV, $v$, being defined through the relation
\begin{eqnarray}
	v^2 = v_1^2 +v_2^2 +v_u^2 +v_d^2 \approx (174 ~{\rm GeV})^2 \,.
\end{eqnarray}
 The diagonal mass matrices can then be
 obtained via the following biunitary transformations:
\begin{subequations}
	\label{e:bidiag}
	\begin{eqnarray}
	D_u &=& U_u   M_u   V_{u}^\dagger = {\rm diag}(m_u,~ m_c,~ m_t ) \,, \\
	D_d &=& U_d   M_d   V_{d}^\dagger = {\rm diag}(m_d,~ m_s,~ m_b ) \,.
	\end{eqnarray}
\end{subequations}
Following this convention for the biunitary transformations, the CKM matrix will be given by
\begin{eqnarray}
\label{e:ckm}
V_{\rm CKM}= U_u  U_d^\dagger \,.
\end{eqnarray}
The matrices $U_u$ and $U_d$ are obtained by diagonalizing $M_u M_u^\dagger$ and $M_d M_d^\dagger$ respectively, which
 can be calculated from Eq. (\ref{e:massmat}) as follows:
\begin{subequations}
	 \begin{eqnarray}
	\label{e:matsqup}
 M_u M_u^\dagger &=& \begin{pmatrix}
             A_u^2 v_1^2 + (B_u^2 + C_u^2) v_2^2 & (-A_u^2 + B_u^2 + C_u^2) v_1 v_2 & (C_u Y_u + B_u X_u) v_2 v_u \\
             (-A_u^2 + B_u^2 + C_u^2) v_1 v_2 & (B_u^2 + C_u^2) v_1^2 + A_u^2 v_2^2 & (C_u Y_u + B_u X_u) v_1 v_u \\
             (C_u Y_u + B_u X_u) v_2 v_u & (C_u Y_u + B_u X_u) v_1 v_u & (Y_u^2 + {X_u}^2) v_u^2    
	\end{pmatrix} \,, \\
	\label{e:matsqdown}
M_d M_d^\dagger &=& \begin{pmatrix}
         A_d^2 v_1^2 + (B_d^2 + C_d^2) v_2^2 & (A_d^2 - B_d^2 - C_d^2) v_1 v_2 & (C_d Y_d + B_d X_d) v_2 v_d \\
            (A_d^2 - B_d^2 - C_d^2) v_1 v_2 & (B_d^2 + C_d^2) v_1^2 + A_d^2 v_2^2 & -(C_d Y_d + B_d X_d) v_1 v_d \\
         	(C_d Y_d + B_d X_d) v_2 v_d & -(C_d Y_d + B_d X_d) v_1 v_d & (Y_d^2 + {X_d}^2) v_d^2
	\end{pmatrix} \,.
	\end{eqnarray}
\end{subequations}
%
%
As a matter of fact, both $M_uM_u^\dagger$ and $M_dM_d^\dagger$
can be fully diagonalized analytically by sequentially operating the following matrices:
\begin{eqnarray}
\label{e:matmudiag}
O_\beta = \begin{pmatrix} 
	                        \cos\beta & -\sin\beta & 0\\
                          	\sin\beta & \cos\beta & 0\\
                          	0 &0 & 1
            	\end{pmatrix}\,, \quad
O_\theta^{u,d}= \begin{pmatrix}
                     1 & 0 & 0\\
            	   0 &	\cos\theta_{u,d} & -\sin\theta_{u,d} \\
            	   0 &	\sin\theta_{u,d} & \cos\theta_{u,d}  		
            	\end{pmatrix}\,,
\end{eqnarray} 
where $\tan\beta=v_2/v_1$ and $\theta_{u,d}$ will be defined shortly.
As a first step, we notice that $M_uM_u^\dagger$ and $M_dM_d^\dagger$ can be block diagonalized using $O_\beta$ as
\begin{subequations}
\label{e:block}
	\begin{eqnarray}
	\label{e:mublock}
	(M_u^2)_{\rm Block} \equiv O_\beta M_u M_u^\dagger O_\beta^\dagger &=& \begin{pmatrix}
	A_u^2 v_{12}^2 & 0 & 0\\
	0 & (B_u^2 + C_u^2) v_{12}^2 & (C_u Y_u + B_u X_u) v_{12} v_u \\
	0 & (C_u Y_u + B_u X_u) v_{12} v_u & (Y_u^2 + {X_u}^2) v_u^2
	\end{pmatrix} , \\
	\label{e:mdblock}
	(M_d^2)_{\rm Block} \equiv O_\beta^\dagger M_dM_d^\dagger O_\beta &=& \begin{pmatrix}
	A_d^2 v_{12}^2 & 0 & 0\\
	0 & (B_d^2 + C_d^2) v_{12}^2 & -(C_d Y_d + B_d X_d) v_{12} v_d\\
	0 & -(C_d Y_d + B_d X_d) v_{12} v_d & (Y_d^2 + {X_d}^2) v_d^2
	\end{pmatrix} ,
	\end{eqnarray}
\end{subequations}
where, as we will see shortly, $v_{12}^2 = v_1^2 + v_2^2$ is the total VEV that is primarily responsible for the light quark masses.
Quite clearly, the remaining $2\times 2$ block in the up and down sectors can be diagonalized using $O_\theta^u$
and $O_\theta^d$, respectively. This second stage of diagonalization allows us to express $\theta_u$ and $\theta_d$ in terms
of the Yukawa couplings and the VEVs as follows:
\begin{subequations}
\label{e:phiud}
\begin{eqnarray}
	\tan 2\theta_u &=& \frac{2(C_u Y_u + B_u X_u) v_{12} v_u}{(Y_u^2 + {X_u}^2) v_u^2- (B_u^2 + C_u^2) v_{12}^2} \,, \\
	\tan 2\theta_d &=& -\frac{2(C_d Y_d + B_d X_d) v_{12} v_d}{(Y_d^2 + {X_d}^2) v_d^2- (B_d^2 + C_d^2) v_{12}^2} \,.
\end{eqnarray}
\end{subequations}
Thus, the full diagonalization in the up and down sectors can be expressed as
\begin{subequations}
\label{e:fulldiag}
\begin{eqnarray}
\label{e:eqmudiag}
D_{u}^2 = O_\theta^u  O_\beta (M_uM_u^\dagger) O_\beta^\dagger {O_\theta^u}^\dagger &\equiv& 
{\rm diag} (m_{u}^2, m_{c}^2, m_{t}^2) \,, \\
\label{e:eqmddiag}
D_{d}^2= O_\theta^d  O_\beta^\dagger  (M_dM_d^\dagger) O_\beta {O_\theta^d}^\dagger &\equiv& 
{\rm diag} (m_{d}^2, m_{s}^2, m_{b}^2)\,.
\end{eqnarray}
\end{subequations}
Following our convention in \Eqn{e:bidiag}, the matrices $U_u$ and $U_d$ can be extracted as follows:
\begin{eqnarray}
\label{e:uu}
U_u = O_\theta^u O_\beta \,, \qquad U_d = O_\theta^d O_\beta^\dagger \,.
\end{eqnarray}
%
Thus from \Eqn{e:ckm}, the CKM matrix is obtained as
{\small
\begin{eqnarray}
\label{e:ckmmat}
V_{\rm CKM}= \begin{pmatrix}
\cos2 \beta & - \cos\theta_d \sin 2\beta & - \sin 2\beta \sin\theta_d\\
\cos\theta_u \sin2 \beta & \cos2 \beta \cos\theta_d \cos\theta_u + \sin\theta_d \sin\theta_u & \cos2 \beta \cos\theta_u \sin\theta_d - 
\cos\theta_d \sin\theta_u\\
\sin2 \beta \sin\theta_u & -\cos\theta_u \sin\theta_d + 
\cos2 \beta \cos\theta_d \sin\theta_u & 
\cos\theta_d \cos\theta_u + \cos 2 \beta \sin\theta_d \sin\theta_u
\end{pmatrix} .
\end{eqnarray}
}
To make the connection between \Eqs{e:ckmmat}{e:wolf} apparent, we assume that
$v_{12}$ is responsible for the masses of the first two generations of quarks whereas $v_u$ and $v_d$ primarily contribute
to the third generation masses in the up and down sector, respectively. Therefore, it is quite natural to expect
$v_{12}\ll v_{u,d}$. From \Eqn{e:block} we identify the first generation quark masses as
\begin{eqnarray}
\label{e:mass1}
	m_u^2 = A_u^2 v_{12}^2 \,, \qquad m_d^2 = A_d^2 v_{12}^2 \,.
\end{eqnarray}
Furthermore, using the VEV hierarchy $v_{u,d}\gg v_{12}$ we can approximate \Eqn{e:phiud} as
\begin{subequations}
	\label{e:phiud1}
	\begin{eqnarray}
	\theta_u &\approx& \frac{(C_u Y_u + B_u X_u)}{(Y_u^2 + {X_u}^2)}\frac{v_{12}}{v_u} \approx \order(\frac{v_{12}}{v_u}) \,, \\
	\theta_d &\approx& -\frac{(C_d Y_d + B_d X_d)}{(Y_d^2 + {X_d}^2)}\frac{v_{12}}{v_d} \approx \order(\frac{v_{12}}{v_d}) \,,
	\end{eqnarray}
\end{subequations}
where we are implicitly assuming that the involved Yukawa couplings have similar orders of magnitude. It is also quite
reasonable to take $v_{12}\sim \order(1~{\rm GeV})$ and $v_{u,d}\sim \order(100~{\rm GeV})$ so that the ratio
$v_{12}/v_{u,d}$ comes out to be $\order(\lambda^2)$. Therefore, from \Eqn{e:phiud1} we conclude
\begin{eqnarray}
\label{e:thetas}
	\sin\theta_{u,d} \approx \order(\lambda^2) \,, \qquad \cos\theta_{u,d} \approx \order(1) \,.
\end{eqnarray}
Moreover, if we identify $\sin2\beta$ as the Cabibbo mixing, namely,
\begin{eqnarray}
\label{e:cab}
 \sin2\beta = \lambda \,,
\end{eqnarray}
then \Eqn{e:ckmmat} resembles exactly to \Eqn{e:wolf}. All these intuitive results will be validated later by providing
explicit numerical benchmarks.

\begin{wrapfigure}{r}{0.5\textwidth}
\centering
\includegraphics[width=0.47\textwidth]{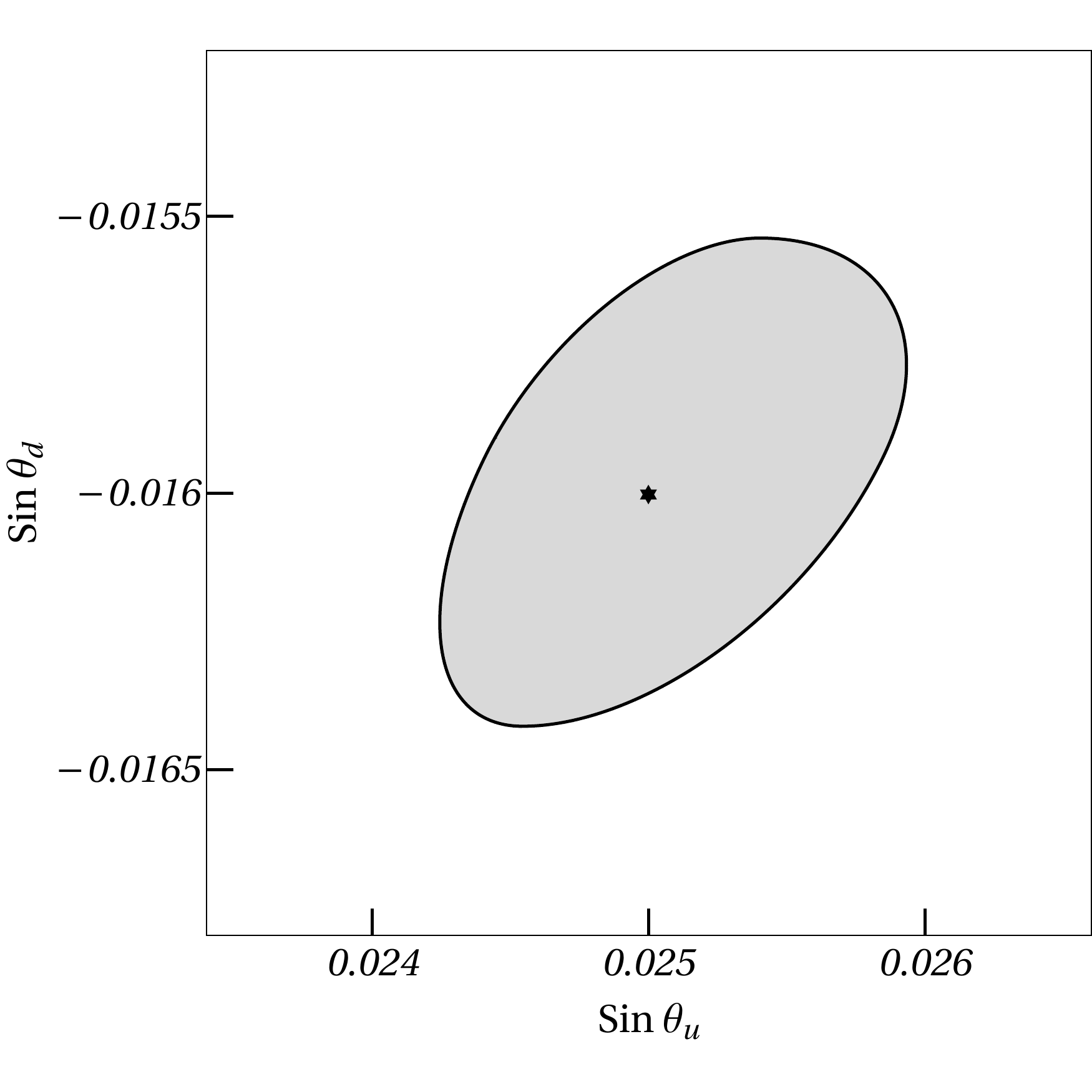}
\caption{\em\small A representative allowed region in the $\sin\theta_u$-~$\sin\theta_d$ plane from the uncertainties in $\theta_{23}$ and $\theta_{13}$. For this plot, we have fixed $\sin2\beta$ at its best-fit value given in \Eqn{e:ckmrange}. The best-fit point 
	in the $\sin\theta_u$-~$\sin\theta_d$ plane	is marked with a star $(\star)$.
	\label{fig:CKM}}
\end{wrapfigure}

Given the structure of the CKM matrix predicted by the model as a function of $\beta$ and $\theta_{u,d}$,   
shown in Eq.~\eqref{e:ckmmat}, it is possible to extract the quark mixing angles by comparing the CKM   
matrix with the standard parametrization\cite{Antusch:2003kp}.  
This, in turn, allows us to find the following best-fit values of $\beta$ and $\theta_{u,d}$ such that the quark mixing angles  
are compatible with the observed values\cite{ParticleDataGroup:2020ssz}:  
\begin{equation}
\label{e:ckmrange}
\sin2\beta \approx 0.2265\,, \qquad \theta_u \approx \pm 0.025, \qquad \theta_d \approx \mp 0.016. 
\end{equation}
As expected, the above values for $\sin 2\beta$ and $\theta_{u,d}$ conform well to our intuitive expectations of Eqs.~\eqref{e:thetas}  
and \eqref{e:cab}.
Fixing $\sin2\beta$ at its best-fit value, in Fig.~\ref{fig:CKM} we display the region in $\sin\theta_u$-~$\sin\theta_d$ plane allowed by
 the experimental uncertainties.
%


%
For the sake of completeness, we also calculate the mass eigenvalues for the second and third generation of quarks
by diagonalizing the $2\times 2$ submatrices in \Eqn{e:block}. In the up quark sector, we can compare the traces to write
\begin{eqnarray}
	m_c^2 +m_t^2 = (B_u^2 + C_u^2) v_{12}^2 + (Y_u^2 + {X_u}^2) v_u^2 \,.
\end{eqnarray}
Keeping in mind the hierarchies, $v_u\gg v_{12}$ and $m_t \gg m_c$, the above relation can be approximated to
express the top quark mass as
\begin{eqnarray}
\label{e:mt}
	m_t^2 \approx (Y_u^2 + {X_u}^2) v_u^2 \,.
\end{eqnarray}
Again, from the determinant of the $2\times 2$ block in \Eqn{e:mublock}, we may write
\begin{eqnarray}
	m_c^2 m_t^2 = (B_uY_u-C_uX_u)^2 v_{12}^2 v_u^2 \,.
\end{eqnarray}
Using the expression for $m_t$ from \Eqn{e:mt}, we can extract the charm quark mass as
\begin{eqnarray}
\label{e:mc}
	m_c^2 \approx \frac{(B_uY_u-C_uX_u)^2}{(Y_u^2 + {X_u}^2)} v_{12}^2 \,.
\end{eqnarray}
Following the same steps in the down sector, we can obtain
\begin{eqnarray}
	\label{e:ms}
	m_s^2 &\approx& \frac{(B_dY_d -C_dX_d)^2}{(Y_d^2 + {X_d}^2)} v_{12}^2 \,, \\
	\label{e:mb}
	m_b^2 &\approx& (Y_d^2 + {X_d}^2) v_d^2 \,.
\end{eqnarray}
At this point, we wish to emphasize that, assuming the Yukawas couplings to be similar for a particular sector,
an obvious outcome of our model is 
\begin{eqnarray}
	\frac{m_c}{m_t} \approx \frac{v_{12}}{v_{u}} \sim \order(\lambda^2) \,, \qquad 
	 \frac{m_s}{m_b} \approx \frac{v_{12}}{v_{d}} \sim \order(\lambda^2) \,,
\end{eqnarray}
which agrees with the observations.

\begin{figure}
 \centering
        \includegraphics[width=0.35\linewidth]{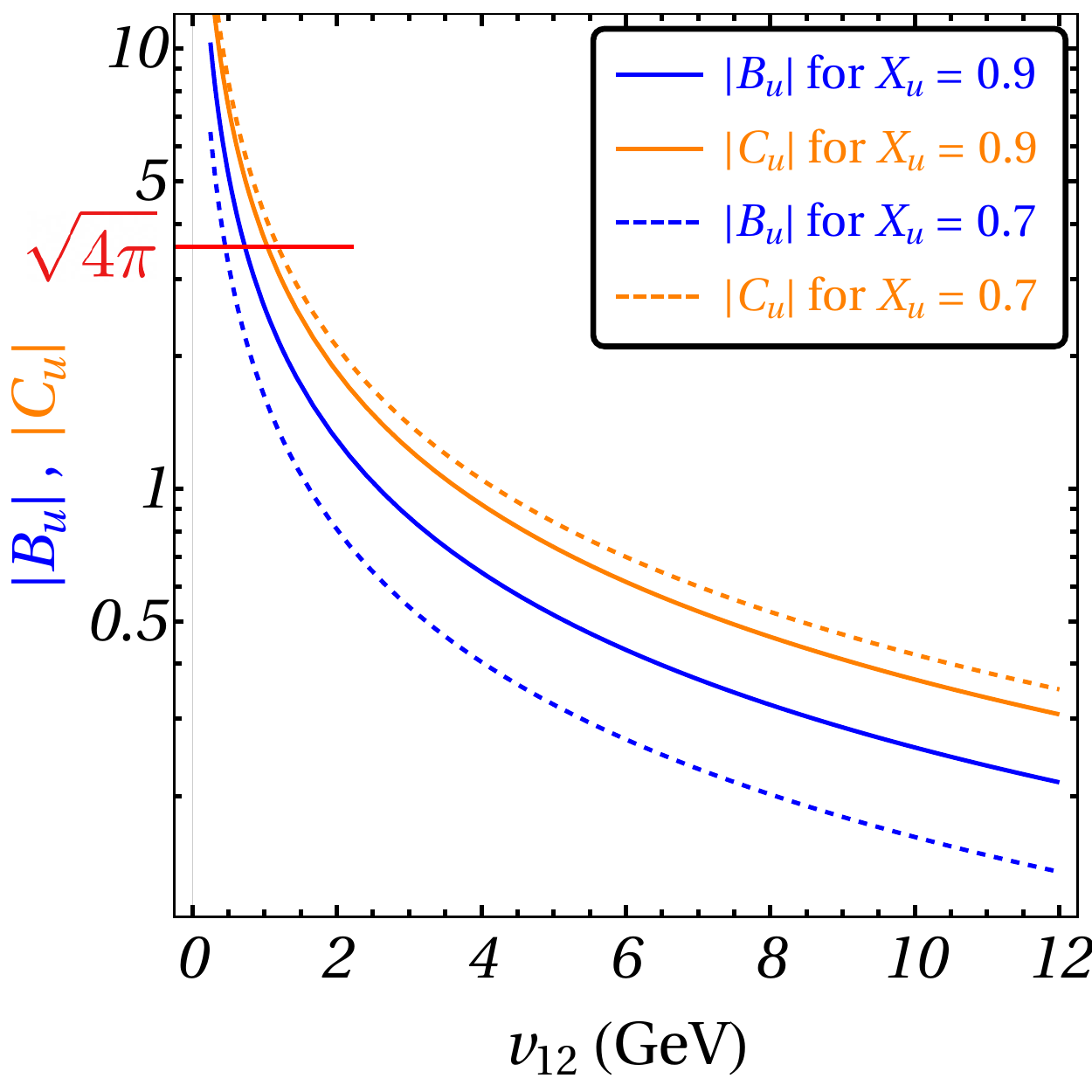}\qquad \qquad \qquad \includegraphics[width=0.35\linewidth]{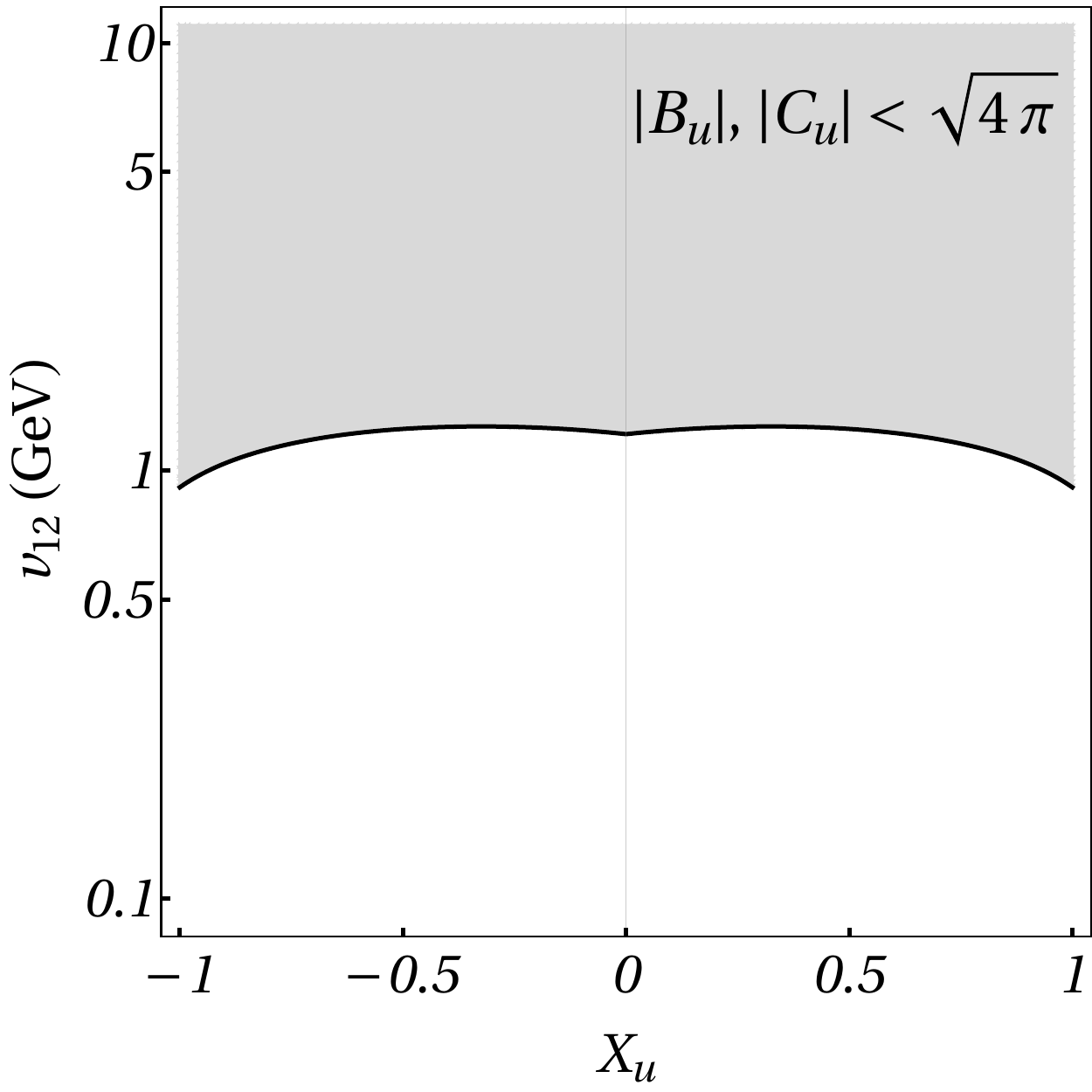}
    	 \caption{\em\small \textbf{Left:} Values of $\lvert B_u\rvert$ (blue) and $\lvert C_u \rvert$ (orange) compatible with the experimental values for the quark masses and mixing angles, for the benchmark points of $ X_u = 0.9$ (solid line) and $X_u = 0.7$ (dashed), as a function of $v_{12}$. The perturbative bound is marked in red. \textbf{Right:} $\lvert B_u \rvert$ and $\lvert C_u\rvert$ are perturbative in the shaded region in the $X_u$- $v_{12}$ plane.
    	  We have assumed $v_u = 150$~GeV and $\sin\theta_u = 0.025$ for both cases.
	 \label{fig:v12}}
\end{figure}

From Eqs.~\eqref{e:mt} and \eqref{e:mc}, we see that the third and second generation masses are controlled by  
$v_u$ and $v_{12}$, respectively. We can wonder how perturbativity may affect the model at hand, since  
$m_t \approx \order(v_u)$ and $m_c \approx \order(v_{12})$ already\cite{Chanowitz:1978mv}. Fig.~\ref{fig:v12} illustrates how arbitrarily low values of $v_{12}$ may jeopardize the perturbativity of the theory. By choosing $v_u = 150$ GeV, and two example values for
$X_u = 0.7$ and  $0.9$, we can see from Fig.~\ref{fig:v12} that to have $B_u$ and $C_u$ in the perturbative regime, we should have $v_{12}\geq \mathcal{O}(1~\text{GeV})$.

Finally, to provide explicit justification to these intuitive
expectations, we consider the following benchmark
\begin{equation}
\label{e:bench}
\begin{matrix}
 v_{12} = {2~{\rm GeV}}, &  v_u ={150~{\rm GeV}}, & v_d \approx {88~{\rm GeV}}, & & \\
 A_u \approx 1.08 \!\!\times \!\! 10^{-3}, & B_u \approx 1.69, & C_u \approx 1.50, & X_u \approx 1.04,  &  Y_u \approx 0.49, \\
A_d \approx 2.34 \!\!\times \!\! 10^{-3}, & B_d \approx 3.65 \!\!\times \!\! 10^{-2},  & C_d \approx 4.41 \!\!\times \!\! 10^{-2},  & X_d \approx 4.73 \!\!\times \!\! 10^{-2},  & Y_d  \approx -3.20 \!\!\times \!\! 10^{-3},
\end{matrix}
\end{equation}
which results in the following values of the quark masses and mixing angles
\begin{subequations}
	\label{e:values}
	\begin{eqnarray}
	&& m_u = {2.2~{\rm MeV}} \,, \qquad m_c ={1.27~{\rm GeV}} \,, \qquad  m_t ={173~{\rm GeV}} \,, \\
	&& m_d = {4.7~{\rm MeV}} \,, \qquad m_s ={0.093~{\rm GeV}} \,, \qquad  m_b ={4.18~{\rm GeV}} \\
	&& \sin\theta_{12} =  0.2265 \,, \qquad \sin\theta_{13} = 0.0036  \,, \qquad  \sin\theta_{23} = 0.041 \,,
	\end{eqnarray}
\end{subequations}
which are in agreement with the corresponding observations\cite{ParticleDataGroup:2020ssz}.

\pagebreak

In passing, let us highlight the most notable outcomes of our model:
\begin{itemize}
	\item The hierarchy of the Yukawa couplings is diluted by two orders of magnitude, at least. Recall that, in
	the SM, $m_t=174$~GeV and $m_{u,d}\sim\order({10^{-3}~\rm GeV})$ imply that the quark Yukawa couplings span
	five orders of magnitudes. We dampen this problem by assuming that the first two generations of quarks
	receive their masses from $v_{12}$ which is of $\order(1~{\rm GeV})$. This means, the first generation
	Yukawas are, at worst, of $\order(10^{-3})$ whereas the second generation Yukawas can be of $\order(1)$.
	This feature is quite evident from the benchmark values given in Eq.~\eqref{e:bench}.
	\item We have introduced $\phi_{u,d}$ dedicated for masses of the third generation of quarks. Quite naturally,
	we expect, $v_{u,d}\sim \order(100~{\rm GeV})$ so that the top-Yukawa is of $\order(1)$. Thus, we should have
	the ratio $v_{12}/v_{u,d}\sim \order(\lambda^2)$. It is very interesting to note that, this automatically
	conforms to $m_2/m_3\approx v_{12}/v_{u,d}\sim \order(\lambda^2)$ where $m_k$ is the mass for the $k$-th
	generation of quark. Quite clearly, this is a natural upshot of our model.
	\item We have connected the quark mixings with the dynamics of the scalar sector. We have shown that the Cabibbo
	part of the quark-mixing stems purely from the ratio $v_2/v_1$ (see Eq.~\eqref{e:cab}). The smallness of the off-Cabibbo elements
	of the CKM matrix is further connected to the VEV hierarchy $v_{12}\ll v_{u,d}$. In other way, we are suggesting
	that the fact that the third generation of quarks are much heavier than the first two generations, is
	intimately connected to the smallness of the off-Cabibbo elements.
\end{itemize}
Finally, our current model is not just all about aesthetics, it will have other observable consequences too.
The fact that the Yukawa Lagrangian of our model contains fewer parameters than that of the SM comes at the cost
of making the scalar potential substantially more involved containing four scalar doublets. This means that
the Higgs boson observed at the LHC is not the only fundamental scalar in nature, it is just the first one
in series of many others to follow. The physical Higgs bosons will emerge from mixings among the four scalar
doublets. Expanding the scalar doublets as
\begin{eqnarray}\label{eq:Phis}
	\phi_k = \begin{pmatrix}
	\varphi_k^+ \\  v_k + (h_k+iz_k)/\sqrt{2}
	\end{pmatrix} \,,
	\qquad k = 1,2, u,d,
\end{eqnarray}
after the spontaneous symmetry breaking, the SM-like Higgs boson, $h$, can be extracted as follows\cite{Das:2019yad}:
\begin{eqnarray}
	h = \frac{1}{v}(v_1h_1 +v_2h_2 +v_uh_u +v_dh_d) \,.
\end{eqnarray}
This particular linear combination of the component fields will mimic the SM Higgs in its tree-level couplings
and will not induce flavor changing neutral currents (FCNCs) at the tree-level. However, the other physical neutral scalars, in
general, will possess tree-level FCNCs which means they have to be quite heavy to evade the experimental constraints.
To have some intuitions on the FCNC couplings, we analyze the matrices, $N_d^{1,2,d}$, which control them in the 
down sector.  
Below, we show approximate expressions for these matrices:
\begin{subequations}
	\label{e:nd}
	\begin{eqnarray}
	N_d^1 &\approx& \frac{1}{\sqrt{2} v_{12}} \begin{pmatrix} m_d \cos\beta & - m_s \sin\beta & m_b \theta_d \sin\beta \\ -m_d \sin_\beta & -m_s \cos\beta & m_b \theta_d \cos\beta \\ -m_d \theta_d \sin\beta & -m_s \theta_d \cos\beta & m_b \theta_d^2 \cos\beta \end{pmatrix}, \\
	N_d^2 &\approx& \frac{1}{\sqrt{2} v_{12}} \begin{pmatrix} m_d \sin\beta & m_s \cos\beta & -m_b \theta_d\cos\beta \\ m_d \cos\beta & -m_s \sin\beta & m_b \theta_d \sin\beta \\ m_b \theta_d \cos\beta & -m_s \theta_d \sin\beta & m_b \theta_d^2 \sin\beta \end{pmatrix}, \\
	N_d^d &\approx& \frac{1}{\sqrt{2} v_{d}} \begin{pmatrix} 0 & 0 & 0 \\ 0 & 0 & -m_b \theta_d \\ 0 & 0 & m_b \end{pmatrix}.
	\end{eqnarray}
\end{subequations}
From the above expressions, we note that the magnitude of the largest off-diagonal
element, for our chosen benchmarks of \Eqs{e:ckmrange}{e:bench}, is $0.033$, which is quite small. On top of this,
the flavor constraints may be further relaxed if we remember the following points:
\begin{itemize}
	\item The actual FCNC matrices that control the couplings of the physical
	neutral scalars are orthogonal linear combinations (dictated by the scalar
	potential) of $N_d^1$, $N_d^2$, $N_d^d$ and $N_d^u$ where $N_d^u=0$ simply
	because $h_u$ does not couple to the down-type quarks.
	\item A cancellation may be arranged between the scalar and pseudoscalar diagrams
	appearing in the FCNC process\cite{Nebot:2015wsa}.
\end{itemize}
%
%
%
%
Furthermore, it should also be noted that low values of the VEVs, especially $v_{12}\sim \order(1~{\rm GeV})$, will not necessarily imply
the existence of light nonstandard scalars if we include terms that softly break the $D_4$ symmetry in
the scalar potential\cite{Bhattacharyya:2014oka,Faro:2020qyp, Carrolo:2021euy} (please refer to the appendix \ref{sec:app} for details).
A detailed study of the scalar potential along with the analysis of the flavor constraints is reserved for
a future work. Nevertheless, our current paper can be considered as a proof-of-concept for a novel idea that it
might be possible to ascribe the quark flavor hierarchies primarily to the hierarchies in the VEVs all of which
add together to constitute the total electroweak VEV\footnote{This is in stark contrast with the Froggatt-Nielsen mechanism \cite{Froggatt:1978nt}, where flavon VEVs are usually much higher than the electroweak scale.}. 
Thus, in other words, we have demonstrated that, to have an insight into the flavor puzzle, it might not be necessary to appeal to flavor symmetry breaking scales much higher than the EW scale. 
Moreover, the fact that such a scheme can be easily
accommodated in a relatively simple and intuitive theoretical set-up, makes our model an interesting addition
to the existing literature on flavor model building.

\appendix
\section{Discussion of the Scalar Potential}
\label{sec:app}

It might be natural to wonder whether the VEV hierarchies and the scalar mass spectrum required for our model to work are admissible by the scalar sector of our model obeying the $D_4$ symmetry.
Here, we will explicitly demonstrate that such features can indeed be achieved when the scalar potential contains terms that softly-break the symmetry.
In view of this, let us write down the scalar potential as follows:
\begin{equation}
V(\phi) = V_{\rm quadratic} + V_{\rm quartic},
\end{equation}
where,
\begin{eqnarray}
\label{D4pot}
V_{\rm quartic} &=& \lambda_1 \left( \phi^\dagger_1 \phi_1+ \phi^\dagger_2 \phi_2 \right)^2 + \lambda_2 \left( \phi^\dagger_1 \phi_2 - \phi^\dagger_2 \phi_1 \right)^2 +  \lambda_3 \left( \phi^\dagger_1 \phi_2 + \phi^\dagger_2 \phi_1 \right)^2 +  \lambda_4 \left( \phi^\dagger_1 \phi_1 - \phi^\dagger_2 \phi_2 \right)^2  \nonumber \\
&&+ \lambda_5 \left( \phi_u^\dagger \phi_u\right)^2 + \lambda_6 \left( \phi_d^\dagger \phi_d\right)^2 + \lambda_7 \left( \phi_1^\dagger \phi_1 + \phi_2^\dagger \phi_2 \right)\left( \phi_d^\dagger \phi_d \right) + \lambda_8 \left( \phi_1^\dagger \phi_1 + \phi_2^\dagger \phi_2 \right)\left( \phi_u^\dagger \phi_u \right)  \nonumber \\
&&+ \lambda_9 \left( \phi_u^\dagger \phi_u \right) \left( \phi_d^\dagger \phi_d \right) + \lambda_{10}\left[ \left( \phi_u^\dagger \phi_1\right) \left( \phi_1^\dagger \phi_u\right) +  \left( \phi_u^\dagger \phi_2\right) \left( \phi_2^\dagger \phi_u\right)\right]  \nonumber \\
&&+  \lambda_{11}\left[ \left( \phi_d^\dagger \phi_1\right) \left( \phi_1^\dagger \phi_d\right) +  \left( \phi_d^\dagger \phi_2\right) \left( \phi_2^\dagger \phi_d\right)\right]   \nonumber \\
&&+ \lambda_{12} \left[ \left( \phi_u^\dagger \phi_1\right)^2 + \left( \phi_u^\dagger \phi_2\right)^2 + \left( \phi_1^\dagger \phi_u\right)^2 + \left( \phi_2^\dagger \phi_u\right)^2\right]  \nonumber \\
&&+ \lambda_{13} \left[ \left( \phi_d^\dagger \phi_1\right)^2 + \left( \phi_d^\dagger \phi_2\right)^2 + \left( \phi_1^\dagger \phi_d\right)^2 + \left( \phi_2^\dagger \phi_d\right)^2\right]  \nonumber \\
&&+ \lambda_{14} \left[ \left( \phi_u^\dagger \phi_1\right) \left( \phi_d^\dagger \phi_1\right) + \left( \phi_u^\dagger \phi_2\right) \left( \phi_d^\dagger \phi_2\right) + \left( \phi_1^\dagger \phi_u\right) \left( \phi_1^\dagger \phi_d\right) + \left( \phi_2^\dagger \phi_u\right) \left( \phi_2^\dagger \phi_d\right)\right]  \nonumber \\
&&+ \lambda_{15} \left[ \left( \phi_u^\dagger \phi_1\right) \left( \phi_1^\dagger \phi_d\right) + \left( \phi_u^\dagger \phi_2\right) \left( \phi_2^\dagger \phi_d\right) + \left( \phi_1^\dagger \phi_u\right) \left( \phi_d^\dagger \phi_1\right) + \left( \phi_2^\dagger \phi_u\right) \left( \phi_d^\dagger \phi_2\right)\right]  \nonumber \\
&&+ \lambda_{16}  \left( \phi_1^\dagger \phi_1 - \phi_2^\dagger \phi_2\right) \left[\left( \phi_u^\dagger \phi_d\right) + \left( \phi_d^\dagger \phi_u \right) \right], \\
V_{\rm quadratic} &=& \mu^2_{11} \left( \phi_1^\dagger \phi_1\right) + \mu^2_{22} \left( \phi_2^\dagger \phi_2\right) + \mu^2_{uu} \left( \phi_u^\dagger \phi_u\right) + \mu^2_{dd} \left( \phi_d^\dagger \phi_d\right)  \nonumber \\
&&+ \mu^2_{12} \left( \phi_1^\dagger \phi_2 + \text{h.c.}\right) + \mu^2_{1u} \left( \phi_1^\dagger \phi_u + \text{h.c.}\right) + \mu^2_{1d} \left( \phi_1^\dagger \phi_d + \text{h.c.}\right)  \nonumber \\
&&+ \mu^2_{2u} \left( \phi_2^\dagger \phi_u + \text{h.c.}\right) + \mu^2_{2d} \left( \phi_2^\dagger \phi_d + \text{h.c.}\right) + \mu^2_{ud} \left( \phi_u^\dagger \phi_d + \text{h.c.}\right) .
\end{eqnarray}
We wish to reiterate that we have softly broken the $D_4$ symmetry by making $V_{\rm quadratic}$ to be completely general, but $V_{\rm quartic}$ obeys the $D_4$ symmetry. 
For simplicity, we take all the parameters to be real. 
Even then, above potential includes a large number of free parameters and its analysis can easily become quite involved. 
However, all we need to show is that (i) the scalar potential does not impose any restrictions on the relative hierarchies of the VEVs, and (ii) the nonstandard scalars may be safely decoupled from the EW scale. Thus, it will suffice if we can show that there exists an intuitive and convenient limit of the D4HDM scalar potential, which can easily accommodate these features. Such a limit can be obtained from \Eqn{D4pot}, by setting $\lambda_{1,5,6} \to \lambda$ together with $\lambda_{7,8,9}\to 2 \lambda$ and the rest of the quartic parameters to zero. The resulting potential reads:
\begin{eqnarray}
\label{MS}
V_{\rm MS} &=&
 \mu_{11}^2 \, \phi_1^\dagger \phi_1 + \mu_{22}^2 \, \phi_2^\dagger \phi_2 + \mu_{uu}^2 \, \phi_u^\dagger \phi_u + \mu_{dd}^2 \, \phi_d^\dagger \phi_d  \nonumber \\
&&+ \left( \mu^2_{12}\,  \phi_1^\dagger \phi_2 + \mu^2_{1u} \, \phi_1^\dagger \phi_u + \mu^2_{1d} \, \phi_1^\dagger \phi_d +
\mu^2_{2u}\,  \phi_2^\dagger \phi_u + \mu^2_{2d}\,  \phi_2^\dagger \phi_d + \mu^2_{ud}\,  \phi_u^\dagger \phi_d + {\rm h.c.} \right)   \nonumber \\
&&+\lambda \left( \phi_1^\dagger \phi_1 + \phi_2^\dagger \phi_2 + \phi_u^\dagger \phi_u + \phi_d^\dagger \phi_d \right)^2,
\end{eqnarray}
which has an enhanced symmetry in the quartic part of the potential. It is quite well-known that the two and three Higgs-doublet versions of such \emph{maximally symmetric} potentials keep the VEV structure general and the nonstandard masses arise solely from the soft-breaking parameters, thereby making them easy to decouple \cite{Darvishi:2019dbh, Faro:2020qyp, Chakraborti:2021bpy}. From the counting of parameters, it is quite foreseeable that such general attributes will still be preserved for the maximally symmetric 4HDM potential displayed in \Eqn{MS}.

To analyze the potential, we decompose the scalar doublets as in \Eqn{eq:Phis}, and the ensuing tadpole equations can be used to express the four diagonal bilinears in terms of the four VEVs as follows
\begin{subequations}
\begin{eqnarray}
\mu_{11}^2 &=& - \frac{1}{2 v_1} \left( 4 \lambda v^2 v_1 + \mu^2_{12} v_2 + \mu^2_{1u}v_u + \mu^2_{1d}v_d \right), \\
\mu_{22}^2 &=& - \frac{1}{2 v_2} \left( 4 \lambda v^2 v_2 + \mu^2_{12} v_1 + \mu^2_{2u}v_u + \mu^2_{1d}v_d \right), \\
\mu_{uu}^2 &=& - \frac{1}{2 v_u} \left( 4 \lambda v^2 v_u + \mu^2_{1u} v_1 + \mu^2_{2u}v_2 + \mu^2_{ud}v_d \right), \\
\mu_{dd}^2 &=& - \frac{1}{2 v_d} \left( 4 \lambda v^2 v_d + \mu^2_{1d} v_1 + \mu^2_{2d}v_u + \mu^2_{ud}v_u \right),
\end{eqnarray}
\end{subequations}
where $v^2 = v_1^2 + v_2^2 + v_u^2 + v_d^2= (174\text{ GeV})^2$ is the EW VEV.
It should be noted that these equations pose no restrictions on the relative magnitudes of the VEVs.
After the spontaneous symmetry breaking, we can extract the scalar mass matrices. Under our assumption of real parameters, the mass matrices of the scalars and pseudoscalars become disentangled.
We now define the mass matrices for the different sectors as follows
\begin{subequations}
\begin{eqnarray}
V^{\rm mass}_S &=& \begin{pmatrix} h_1, & h_2, &h_u, &h_d \end{pmatrix} \frac{\mathcal{M}_S^2}{2} \begin{pmatrix} h_1 \\ h_2 \\ h_u \\ h_d \end{pmatrix}, \\
V^{\rm mass}_P &=& \begin{pmatrix} z_1, & z_2, &z_u, &z_d \end{pmatrix} \frac{\mathcal{M}_P^2}{2} \begin{pmatrix} z_1 \\ z_2 \\ z_u \\ z_d \end{pmatrix}, \\
V^{\rm mass}_C &=& \begin{pmatrix} \varphi^+_1, & \varphi^+_2, &\varphi^+_u, &\varphi^+_d \end{pmatrix} \mathcal{M}_C^2 \begin{pmatrix} \varphi^-_1 \\ \varphi^-_2 \\ \varphi^-_u \\ \varphi^-_d \end{pmatrix}.
\end{eqnarray}
\end{subequations}

Due to the particularly simple structure of the maximally symmetric potential, we find
\begin{eqnarray}
\!\!\!\!\!\! \mathcal{M}^2_P &=& \begin{pmatrix} -\frac{\mu^2_{12} v_2 + \mu^2_{1u} v_u + \mu^2_{1d} v_d }{2 v_1} & \frac{\mu^2_{12}}{2} & \frac{\mu^2_{1u}}{2} & \frac{\mu^2_{1d}}{2} \\
. & -\frac{\mu^2_{12} v_1 + \mu^2_{2u} v_u + \mu^2_{2d} v_d }{2 v_2} & \frac{\mu^2_{2u}}{2} & \frac{\mu^2_{2d}}{2} \\
. & . & -\frac{\mu^2_{1u} v_1 + \mu^2_{2u} v_2 + \mu^2_{ud} v_d }{2 v_u} & \frac{\mu^2_{ud}}{2} \\
. & . & . & -\frac{\mu^2_{1d} v_1 + \mu^2_{2d} v_2 + \mu^2_{ud} v_u }{2v_d} \end{pmatrix}\\
\!\!\!\!\!\! \mathcal{M}^2_S &=& \mathcal{M}^2_P + \begin{pmatrix}  4 \lambda v_1^2 &  4 \lambda v_1 v_2 & 4 \lambda v_1 v_u & 4 \lambda v_1 v_d \\
. & 4 \lambda v_2^2 & 4 \lambda v_2 v_u & 4 \lambda v_2 v_d \\
. & . & 4 \lambda v_u^2 & 4 \lambda v_u v_d \\
. & . & . & 4 \lambda v_d^2 \end{pmatrix}  , \qquad \mathcal{M}^2_C = \mathcal{M}^2_P\, ,
\end{eqnarray}
where we have not shown the lower triangles of the matrices explicitly, because they are symmetric.

We will now parametrize four VEVs following a straightforward generalization of the $\tan\beta$ parameter for the 2HDM. Since all the scalar VEVs contribute to the EW VEV, we define:
\begin{equation}
v_1 = v \cos\beta \cos\beta_1 \cos \beta_2 , \quad v_2 = v \sin\beta \cos\beta_1 \cos\beta_2, \quad v_u = v \sin\beta_1 \cos\beta_2, \quad v_d = v \sin\beta_2, 
\end{equation}
which satisfies $v_1^2 +v_2^2+v_u^2 +v_d^2 = v^2$ by design, and $\tan\beta=v_2/v_1$ still holds. We can rotate to the Higgs basis by shifting all of the VEVs into a single scalar, which will have SM-like couplings to the fermions and gauge bosons,
\begin{equation}
h = \frac{1}{v} \left( v_1 h_1 + v_2 h_2 + v_u h_u + v_d h_d\right).
\end{equation}
As we will see shortly, this state $h$ automatically emerges as a physical mass eigenstate, thus ensuring the compatibility of our model with the measurements of the Higgs signal-strengths.
A straightforward way to rotate to the Higgs basis will be to use the following orthogonal matrix
\begin{equation}
O_\beta = 
\begin{pmatrix}
\cos\beta\cos\beta_1\cos\beta_2 & \sin\beta\cos\beta_1\cos\beta_2 & \sin\beta_1 \cos\beta_2 & \sin\beta_2 \\
-\sin\beta & \cos\beta & 0 & 0 \\
-\cos\beta\sin\beta_1 & -\sin\beta\sin\beta_1 & \cos\beta_1 & 0 \\
-\cos\beta\cos\beta_1\sin\beta_2 & -\sin\beta\cos\beta_1\sin\beta_2 & -\sin\beta_1\sin\beta_2 & \cos\beta_2
\end{pmatrix},
\end{equation}
to obtain the mass matrices in the Higgs basis, as follows:
\begin{subequations}
\begin{eqnarray}
\label{eq13a}
O_\beta \cdot \mathcal{M}^2_S \cdot O_\beta^T&=& \begin{pmatrix} 4 \lambda v^2 & 0\\ 0 & \left(\mathcal{M}^2\right)_{3 \times 3} \end{pmatrix}, \\
O_\beta \cdot \mathcal{M}^2_P \cdot O_\beta^T &=& \begin{pmatrix} 0 & 0\\0 & \left(\mathcal{M}^2\right)_{3 \times 3} \end{pmatrix}, \\
O_\beta \cdot \mathcal{M}^2_C \cdot O_\beta^T &=& \begin{pmatrix} 0 & 0\\0 & \left(\mathcal{M}^2\right)_{3 \times 3} \end{pmatrix}.
\end{eqnarray}
\end{subequations}
Indeed, the Higgs basis automatically block-diagonalizes the mass matrices, especially $\mathcal{M}_S^2$. Thus, from \Eqn{eq13a}, we see that the scalar which features SM-like couplings to the fermions and gauge bosons is also a mass eigenstate of the theory, with squared-mass $m_h^2=4\lambda v^2$. In other words, in this case we are automatically in the alignment limit, and the model comfortably accommodates an SM-like Higgs. Most importantly, the remaining $3\times3$ block matrices related to the nonstandard scalars are identical in the three sectors, and are independent of the EW scale, as they do not feature any dependence on $v$:
\begin{subequations}
\begin{eqnarray}
\left(\mathcal{M}^{2}\right)_{11} &=&  -\frac{1}{2} \tan\beta \left[ \sin\beta \left( \mu_{12}^2 \sin\beta + \mu^2_{1d} \secant\beta_1\tan\beta_2 + \mu^2_{1u} \tan\beta_1 \right) \right.\nonumber \\
&& \left. \quad  + \cos\beta \cot^2\beta \left( \mu^2_{2d} \secant\beta_1 \tan\beta_2 + \mu^2_{2u} \tan\beta_1\right) + \mu^2_{12} \cos^2\beta \left( \cot^2\beta +2\right) \right], \\
\left(\mathcal{M}^{2}\right)_{12} &=&  \frac{1}{2} \left[ \cos\beta \left( \mu^2_{2d} \tan\beta_1 \tan\beta_2 + \mu^2_{2u} \secant\beta_1 \right) - \sin\beta \left( \mu^2_{1d} \tan\beta_1 \tan\beta_2 + \mu^2_{1u} \secant\beta_1 \right) \right], \\ 
\left(\mathcal{M}^{2}\right)_{13} &=&  \frac{1}{2} \secant\beta_2 \left( \mu^2_{2d} \cos\beta - \mu^2_{1d} \sin\beta \right), \\
\left(\mathcal{M}^{2}\right)_{22} &=& -\frac{1}{2} \cosec\beta_1 \left\{ \cos^2\beta_1 \tan\beta_2 \left[ \tan^3\beta_1 \left( \mu^2_{1d} \cos\beta + \mu^2_{2d} \sin\beta \right) + \mu^2_{ud} \right] \right. \nonumber \\ 
&& \quad \left. + \secant\beta_1 \left( \mu^2_{1u} \cos\beta + \mu^2_{ud} \sin\beta \right) \right\}, \\
\left(\mathcal{M}^{2}\right)_{23} &=&  \frac{1}{2} \secant\beta_2 \left[ \mu^2_{ud} \cos\beta_1 - \sin\beta_1 \left( \mu^2_{1d} \cos\beta + \mu^2_{2d} \sin\beta\right)\right], \\
\left(\mathcal{M}^{2}\right)_{33} &=& -\frac{1}{2} \cosec\beta_2 \secant\beta_2 \left( \mu^2_{1d} \cos\beta \cos\beta_1 + \mu^2_{2d} \sin\beta \cos\beta_1 + \mu^2_{ud} \sin\beta_1 \right).
\end{eqnarray}
\end{subequations}
Since these $3\times3$ matrices are identical in all sectors, it is clear that the maximally symmetric case will feature a tier-wise degeneracy, with 3 nonstandard mass scales, implying $M_{H_i} = M_{A_i} = M_{H^\pm_i}=M_i$, where $H_i, A_i, H^\pm_i$ denote the $i$-th nonstandard scalar, pseudoscalar, and charged mass eigenstates, respectively. As these are disentangled from the EW scale, this limit features not only the automatic alignment limit, but also the decoupling limit, and the nonstandard masses can be taken to be as heavy as desired, without any bounds arising from unitarity. This is a consequence of having a single quartic parameter, $\lambda$, which will correspond to the SM quartic coupling.

To provide more detail, we note that $\mathcal{M}^2$ is a real, symmetric $3\times3$ matrix. Thus, it can be diagonalized by an orthogonal matrix with 3 rotation angles:
\begin{equation}
O_\alpha = \begin{pmatrix} 1 & 0 & 0 \\ 0 & \cos\alpha_3 & \sin\alpha_3 \\ 0 & -\sin\alpha_3 & \cos\alpha_3 \end{pmatrix} \cdot
\begin{pmatrix} \cos\alpha_2 & 0 & \sin\alpha_2 \\ 0 & 1 & 0 \\ -\sin\alpha_2 & 0 & \cos\alpha_2 \end{pmatrix} \cdot
\begin{pmatrix} \cos\alpha_1 & \sin\alpha_1 & 0 \\ -\sin\alpha_1 & \cos\alpha_1 & 0 \\ 0 & 0 & 1 \end{pmatrix},
\end{equation}
such that
\begin{equation}
O_\alpha \cdot \left(\mathcal{M}^2\right)_{3\times3}  \cdot O_\alpha^T = \text{diag}\left( M^2_1, M^2_2, M^2_3 \right).
\end{equation}
The above equation will lead to 6 relations between the 6 physical parameters (the 3 nonstandard mass scales, $M_1, M_2, M_3$, and the 3 mixing angles, $\alpha_1, \alpha_2, \alpha_3$ present in $O_\alpha$), and the elements of $\mathcal{M}^2$.
Note that, the elements of $\mathcal{M}^2$ are still independent because of the presence of the 6 off-diagonal soft-breaking parameters. Thus it is possible to adjust these parameters such that we find any desired nonstandard mixings and masses.

To sum up, we find that for the maximally symmetric case, in the presence of soft-breaking terms:
\begin{itemize}
	\item The alignment limit emerges automatically (there is a mass eigenstate which is SM-like).
	\item The nonstandard masses are disentangled from the EW scale (the nonstandard masses are tier-wise degenerate and decoupled).
	\item There is no restriction on the relative hierarchies of the VEVs.
	\item Unitarity and boundedness from below constraints are trivially satisfied, because there is only one quartic parameter $\lambda$, which is related to the SM-like Higgs mass as $m_h^2 = 4 \lambda v^2$.
	\item The $\rho$-parameter is trivially satisfied by the tier-wise degeneracy of the masses \cite{Grimus:2007if, Grimus:2008nb}.
\end{itemize}
Therefore, it should be evident that the maximally symmetric case, in the presence of soft-breaking terms, is phenomenologically viable.
Now, since the more general $D_4$ potential of \Eqn{D4pot} encompasses the maximally symmetric limit of \Eqn{MS}, then there is a finite range of parameter space where these same conclusions apply to the softly-broken $D_4$ case.

\paragraph*{Acknowledgements:}

We thank Palash B. Pal for useful insights about the model. 
We thank Ipsita Saha for help with Mathematica. 
DD thanks the Science and Engineering Research Board, India for financial support
through grant no. SRG/2020/000006.
The work of ML is funded by
Funda\c{c}\~{a}o para a Ci\^{e}ncia e Tecnologia-FCT through Grant
No.PD/BD/150488/2019, in the framework of the Doctoral Programme
IDPASC-PT, and in part by the projects CFTP-FCT Unit 777 
(UIDB/00777/2020 and UIDP/00777/2020), and CERN/FIS-PAR/0008/2019.


\bibliographystyle{JHEP} 
\bibliography{D4.bib}
\end{document}